\documentclass[12pt,preprint]{aastex}

\slugcomment{}

\shorttitle{Synchrotron Self-Compton Model for PKS 2155--304}
\shortauthors{Kusunose, \& Takahara}

\begin{document}
\newcommand{\D}{\displaystyle}

\title{Synchrotron Self-Compton Model for PKS 2155--304}

\author{Masaaki Kusunose}
\affil{Department of Physics, School of Science and Technology,
Kwansei Gakuin University, Sanda 669-1337, Japan}
\email{kusunose@kwansei.ac.jp}

\and

\author{Fumio Takahara}
\affil{Department of Earth and Space Science,
Graduate School of Science, Osaka University,
Toyonaka 560-0043, Japan}
\email{takahara@vega.ess.sci.osaka-u.ac.jp}


\begin{abstract}
H.E.S.S. observed TeV blazar PKS 2155--304 in a strong flare state in 2006 July.
The TeV flux varied on timescale as short as a few minutes,
which sets strong constraints on the properties of the emission region.
By use of the synchrotron self-Compton model, 
we found that models with the bulk Lorentz factor $\sim 100$, 
the size of the emission region $\sim 10^{15}$ cm,
and magnetic field $\sim 0.1$ G explain the observed spectral energy distribution
and the flare timescale $\sim$ a few minutes.
This model with a large value of $\Gamma$ accounts for the emission spectrum
not only in the TeV band but also in the X-ray band.
The major cooling process of electrons/positrons in the jet is inverse Compton scattering 
off synchrotron photons.
The energy content of the jet is highly dominated by particle kinetic energy 
over magnetic energy.
\end{abstract}


\section{Introduction}

In active galactic nuclei various physical processes such as 
particle acceleration, emission of high energy photons, and so on, take place.
Accretion of matter onto the central black holes is thought
to be the energy source of these processes.
When a fraction of matter is ejected from the accretion disks,
relativistic jets of plasmas are formed.
The relativistic jets are thought to explain intense and variable emission from blazars.
The beaming effect amplifies the radiation from jets depending on the opening angle of 
the jets, bulk Lorentz factor, $\Gamma$, and the angle between the line of sight and 
the jet axis, $\theta$.
Some blazars are known to emit very high energy $\gamma$-rays in the TeV energy band.
They are, for example, PKS 2155--304, Mrk 421, and Mrk 501
\citep[e.g.,][]{wag07}.

Recent observations of TeV $\gamma$-rays have found remarkably short time variability 
of TeV blazars. 
In 2006 July PKS 2155--304 with redshift $z = 0.116$ showed an outburst of TeV $\gamma$-rays. 
The average flux during the outburst was more than 10 times typical values observed from 
the object \citep{aha-flare07}.  During this period X-rays were also monitored by {\it Swift}.
In the 0.3--10 keV energy band the X-ray flux increased by a factor of 5 \citep{fos07}.
The observations by the High Energy Stereoscopic System (H.E.S.S.) 
report that the timescale of variation is 
a few minutes \citep{aha-flare07}.  
The well-resolved burst of the TeV $\gamma$-ray flux from PKS 2155--304 
varied on timescales only $\sim 200$ s.
Recently MAGIC also observed short time variation in Mrk 501 \citep{alb07}.
When the size of the emission region in the comoving frame of the jet is denoted by $R$, 
the observed timescale of the variability sets a limit on $R$, i.e.,
$R \lesssim c t_\mathrm{var} {\cal D}/(1+z)$, where $c$ is the speed of light
and ${\cal D} = [\Gamma (1- \beta \cos \theta)]^{-1}$ is 
the beaming factor with $\beta = (1-1/\Gamma^2)^{1/2}$.
\cite{aha-flare07} argued that ${\cal D} \gtrsim$ 60 - 120 $R/R_\mathrm{S}$ is required
to explain $t_\mathrm{var} \sim 2$ min,
where $R_\mathrm{S} = 2 G M/c^2$ is the Schwarzschild radius with 
$G$ and $M \sim 10^9 M_\sun$ being the gravitational constant 
and the central black hole mass, respectively.
Such large values of $\cal D$ are also suggested by \cite{bgl08} recently,
based on the requirements of radiative cooling time and 
optical depth for $\gamma$-rays due to electron-positron pair production.

The emission spectra of blazars are characterized by two peaks in 
the $\nu$-$\nu F_\nu$ representation, where $F_\nu$ is the differential energy flux.
The lower energy peak is located in the optical -- X-ray bands and the higher energy peak is
located in very high energy $\gamma$-ray bands.
The lower energy peak is most probably by synchrotron radiation of nonthermal 
electrons/positrons.
Radiation mechanisms of high energy $\gamma$-rays are thought to be inverse Compton scattering
off soft photons in the leptonic models \citep[e.g.,][]{mar92}
and hadronic interaction of relativistic particles
and photons in the hadronic models \citep[e.g.,][]{mb92,mu03}.
The soft photons of the leptonic models are supplied by synchrotron radiation by relativistic
electrons/positrons in the jet (synchrotron self-Compton model)
or by the external sources such as accretion disks \citep{ds93}
and disk radiation scattered around the jet \citep{sikora94}.
These emission models assume that the emission region is one zone.

Observations of TeV $\gamma$-rays of blazars have revealed that large values of 
${\cal D} \gtrsim 10$ are required to fit the emission spectra of very high energy
$\gamma$-rays by inverse Compton scattering \citep[e.g.,][]{ktk02}.
On the other hand, the observations of the apparent velocity of VLBI knots show
that parsec-scale jets are subrelativistic or at most mildly relativistic 
\citep[][]{pe04}.
To reconcile the discrepancy between 
these values of $\Gamma$, the deceleration of jets is considered.
\cite{gk03} assumed that TeV $\gamma$-rays are produced by inverse Compton scattering off
synchrotron photons that are emitted by decelerated jet components.
\cite{gtc05}, on the other hand, proposed a spine-layer model.
In this model they assumed that a fast moving emission region is surrounded by a slow moving sheath.
Gamma-rays are then produced by inverse Compton scattering off the photons emitted in the spine 
and the sheath.
Although various emission models of jets with multiple radiation zones have been 
proposed as mentioned above,
a simple synchrotron self-Compton (SSC) model is still worth use in obtaining physical 
parameters of the emission regions in jets.

In this paper we show that a simple SSC model can explain the TeV $\gamma$-rays 
and X-rays of PKS 2155--304.
PKS 2155--304 is an interesting source because of its strong TeV $\gamma$-ray 
emission and the short time variability of TeV emission.
Since the redshift of PKS 2155--304 is 0.116, 
the absorption of TeV $\gamma$-rays by extragalactic background light (EBL) is effective 
\citep[e.g.,][]{sms06}.  
This is an ideal object to test EBL models as well as emission models.

In \S \ref{sec:model} we describe the parameters of our numerical calculations
and in \S \ref{sec:results} the values of the parameters are determined by
fitting the observed data.  Finally discussion is given in \S \ref{sec:sum}.


\section{Model Parameters} \label{sec:model}

The emission region is assumed to be a sphere with radius $R$ moving relativistically with
bulk Lorentz factor $\Gamma$.  Below we assume ${\cal D} = \Gamma$,
i.e., $\theta \sim 1/\Gamma$.
We solve kinetic equations of electrons and photons in the emission region \citep{ktl00}.
Here we assume that the plasmas and radiation in the emission region are
in a steady state.
Electrons are continuously injected in the emission region at rate $q_\mathrm{inj}$.
The injection spectrum is given by
\begin{equation}
q(\gamma) = K \gamma^{-p} \exp(-\gamma/\gamma_\mathrm{max}) , 
\quad \gamma > \gamma_\mathrm{min}
\end{equation}
where $p$, $\gamma_\mathrm{min}$, and $\gamma_\mathrm{max}$ are parameters,
and $K$ is the normalization constant determined by
\begin{equation}
q_\mathrm{inj} = \int_{\gamma_\mathrm{min}}^\infty q(\gamma) d \gamma  .
\end{equation}
The electrons escape from the emission region by advection on timescale $f_\mathrm{ad} R/c$,
where $f_\mathrm{ad}$ is a dimensionless parameter.
The cooling processes of the electrons are synchrotron radiation and inverse Compton scattering.
We assume that the emission region has randomly oriented magnetic field $B$.
Photons are assumed to escape from the emission region on timescale $R/c$.
We also include the absorption of $\gamma$-rays in the emission region by $e^\pm$ production
due to photon-photon collisions.
This effect is found to be negligible for $\gamma$-rays with observed energy less than 10 TeV
in our numerical results.

Parameters in our model are $\Gamma$, $R$, $B$, $q_\mathrm{inj}$, $p$, $\gamma_\mathrm{min}$,
$\gamma_\mathrm{max}$, and $f_\mathrm{ad}$.
The dependence of solutions on $f_\mathrm{ad}$ is weak and we set $f_\mathrm{ad} = 4$.
In addition to the above parameters, the cosmological parameters such as 
Hubble constant $H_0$ and the 
density parameters of matter $\Omega_m$ and cosmological constant $\Omega_\Lambda$ are needed
to calculate the luminosity distance and the optical depth for $\gamma$-ray absorption by EBL.
We assume $H_0 = 71$ km s$^{-1}$ Mpc$^{-1}$, $\Omega_m = 0.27$, and $\Omega_\Lambda = 0.73$.

It is known that TeV $\gamma$-rays emitted by distant sources are absorbed by EBL.  
This was first pointed out by \cite{nik62}
and detailed calculations were performed by \cite{gs66} and \cite{jelly66}.
Later \cite{sjs92} proposed that the EBL spectrum is estimated by considering the absorption of
TeV $\gamma$-rays from blazars. 
Since then various models of EBL have been proposed \citep[see][for review]{sms06}.
The redshift of PKS 2155--304 is 0.116 and the optical depth of TeV $\gamma$-rays is
greater than unity for $\gamma$-rays with energy greater than 1 TeV.
In our previous work \citep{kkt06}, we used models by \cite{dk05} to fit the emission
spectrum from H1426+428.  There we found that model LLL of \cite{dk05} is applicable
in the SSC model.
In this paper we use the same EBL model to calculate the deabsorbed TeV spectrum.


\section{Results} \label{sec:results}

\begin{figure}
\includegraphics[angle=0,scale=.59]{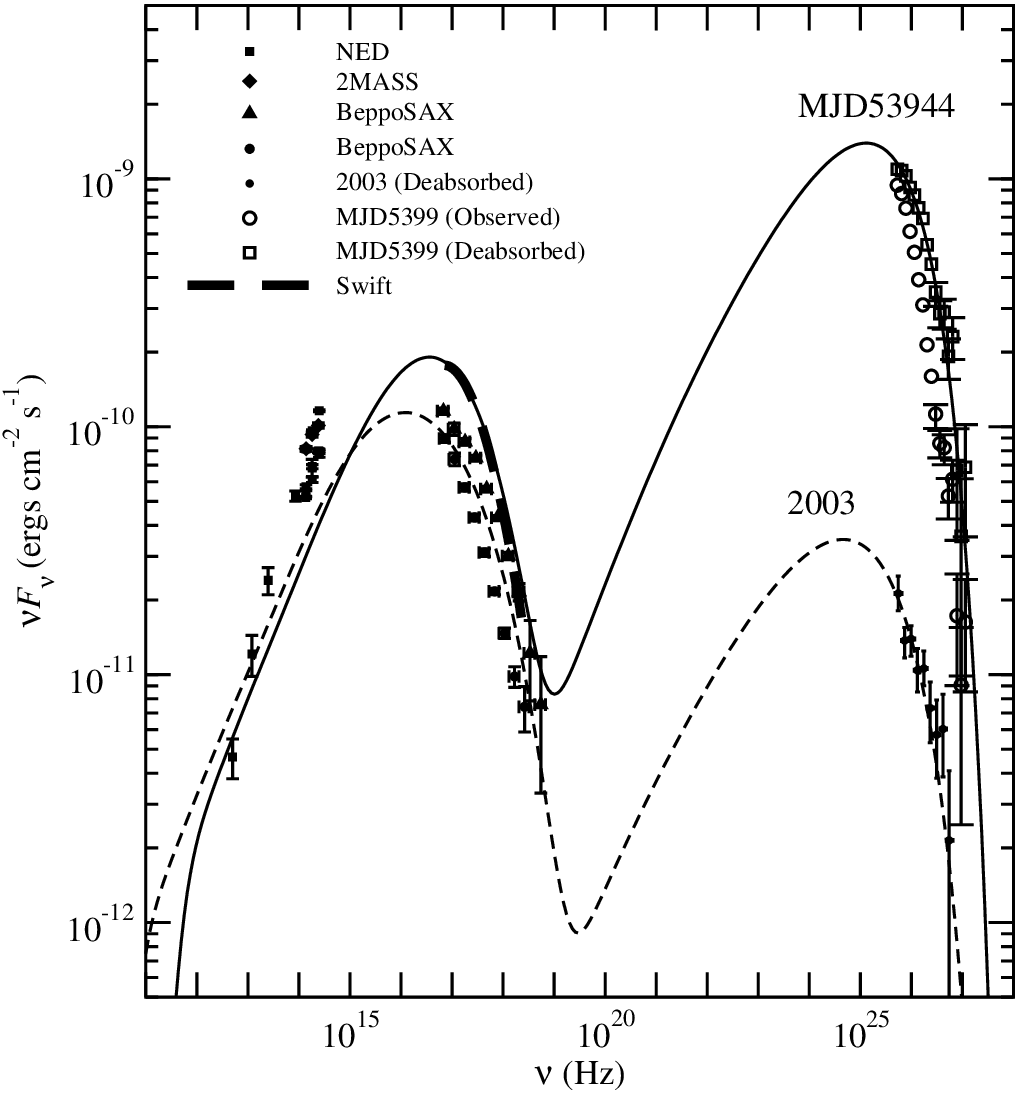}
\caption{The emission spectra of PKS 2155--304.  The TeV data are observed in 2006 July 
(MJD53944) and 2003 July.
The data of MJD53944 (open squares) and those of 2003 July are corrected for absorption by EBL.
The thick dashed line is X-ray data for 2006 July observed by {\it Swift}.
Other X-ray data are obtained by {\it BeppoSAX}, which are not simultaneous with the TeV data.
Model A is shown by a solid line and a model for 2003 is shown by a dashed line.}
\label{fig:photonspec1}
\end{figure}

H.E.S.S. observed a $\gamma$-ray outburst from PKS 2155--304 on 2006 July 28 (MJD 53944) 
\citep{aha-flare07}.
Almost simultaneous observation by {\it Swift} was performed in 
the X-ray band on July 29/30 \citep{fos07}.
The observed emission spectra are shown in Figure \ref{fig:photonspec1}.
The TeV data in 2006 July are from H.E.S.S. \citep{aha-flare07}.
The TeV spectrum is corrected with an EBL model, LLL, of \cite{dk05} for
absorption by $e^\pm$ production through photon-photon collisions.
The X-ray spectrum of 2006 July 29 and 30 is shown by a thick dashed line. 
This spectrum is a log-parabolic model fit given by \cite{fos-ast07}.
Other X-ray, optical, and radio data are not simultaneous with the TeV data.
The X-ray data except those of {\it Swift} are from {\it BeppoSAX} and radio data are from NED.  
2MASS data are also plotted.
The TeV emission spectrum observed in 2003 July \citep{aha-quiet04} is shown 
for comparison.

\begin{figure}
\includegraphics[angle=0,scale=.59]{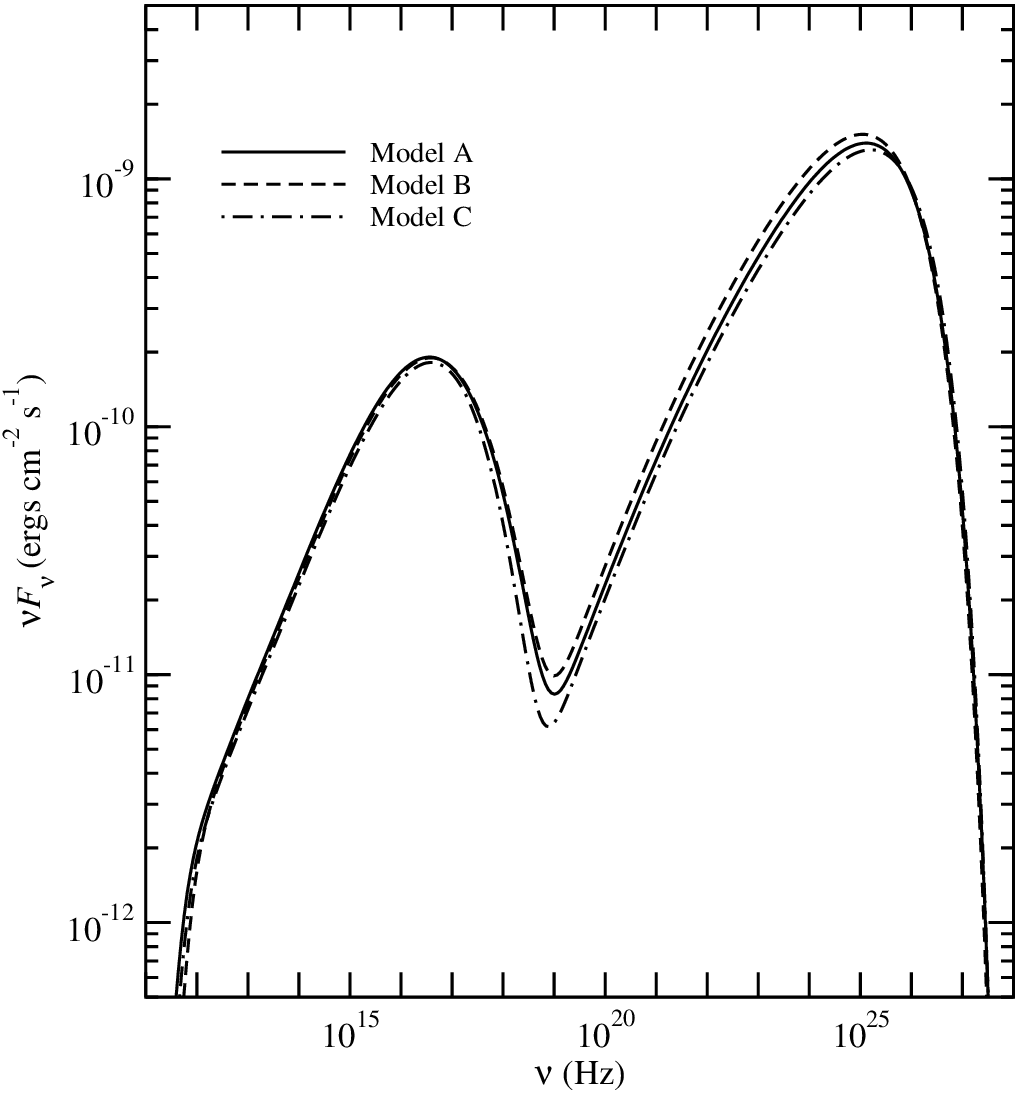}
\caption{SEDs of Models A, B, and C are shown by solid, dashed, and dot-dashed lines,
respectively. }
\label{fig:photon-abc}
\end{figure}

Our models are shown in Figure \ref{fig:photonspec1} by solid and dashed lines for
2006 July and 2003 July, respectively.
We did not fit the emission spectra below $10^{15}$ Hz, 
assuming this emission is from different regions, possibly from extended regions
far away from the central black hole.
The values of the parameters for the flare in 2006 July are the following:
$\Gamma = {\cal D} = 90$, 
$B = 0.1$ G, $R = 9.6 \times 10^{14}$ cm, $p = 1.9$, $\gamma_\mathrm{min} = 10$,
and $\gamma_\mathrm{max} = 8 \times 10^4$ (Model A in Table \ref{tbl-1}).
The value of $R$ is $\sim 3 R_\mathrm{S}$, if $M = 10^9 M_\sun$.
From these parameters, the timescale of variability is 
$t_\mathrm{var} \sim (1+z)R/(c \Gamma) \sim 400$ s.
We also obtained the parameter values which give $t_\mathrm{var} \sim 200$ s.
Those are shown in Table \ref{tbl-1} (Models B and C).
When $t_\mathrm{var} \sim 200$ s is assumed,
the spectral energy distribution (SED) of 2006 July is fitted well
if $\Gamma$ is in the range $100 \lesssim \Gamma \lesssim 150$.
When $\Gamma \lesssim 90$, the fluxes in the tails of the lower and higher peaks of the SED
are too high.  On the other hand, when $\Gamma > 150$, the tail of the lower peak (the X-ray band)
is too steep.
SEDs for Models A, B, and C are shown in Figure \ref{fig:photon-abc}.
In Table \ref{tbl-1} the parameters given in other papers such as
\cite{fos07} and \cite{bgl08} are also listed.
Note that $t_\mathrm{var} = 1$ h is assumed in \cite{fos07}
and that \cite{bgl08} did not perform the spectral fitting.

\begin{figure}
\includegraphics[angle=0,scale=.59]{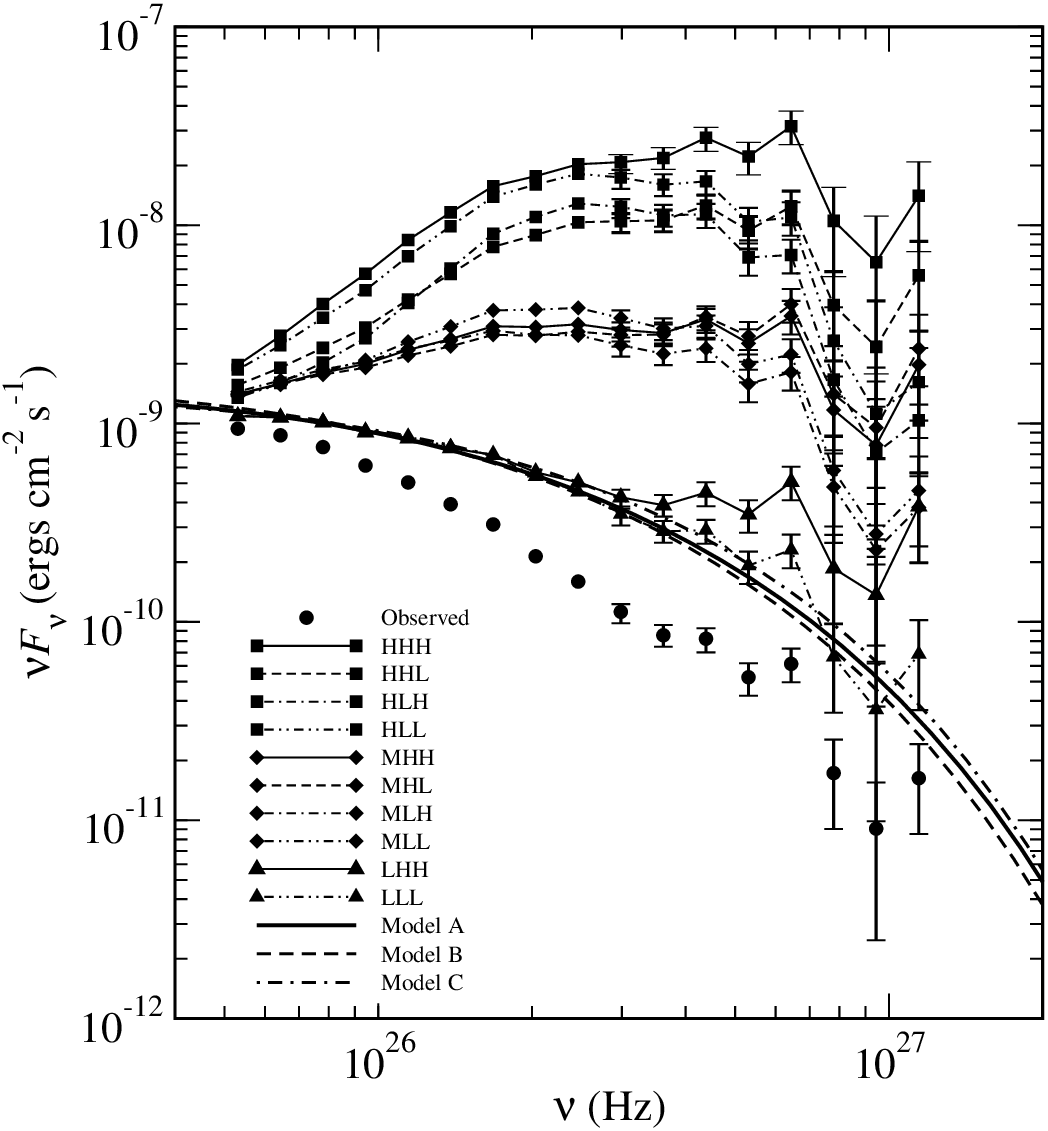}
\caption{The TeV spectrum of 2006 July (filled circles) is deabsorbed with different EBL models
\citep[][]{dk05}. Models A, B, and C are shown by solid, dashed, and dot-dashed lines,
respectively. }
\label{fig:photon-tev}
\end{figure}

In Figure \ref{fig:photon-tev} various deabsorbed TeV spectra are compared with our models.
The deabsorbed spectra are calculated with different EBL models given by \cite{dk05}.
When EBL models other than LLL, LHH, LLH, and LHL are applied
(LLH and LHL are not shown in the figure),
the TeV spectrum has a peak at $\sim 3 \times 10^{26}$ Hz.
Such a spectrum is difficult to produce by the one-zone SSC model,
if the X-rays are emitted in the same region as TeV $\gamma$-rays.
In particular the Klein-Nishina effect suppresses the emission in the TeV band.

\begin{figure}
\includegraphics[angle=0,scale=.59]{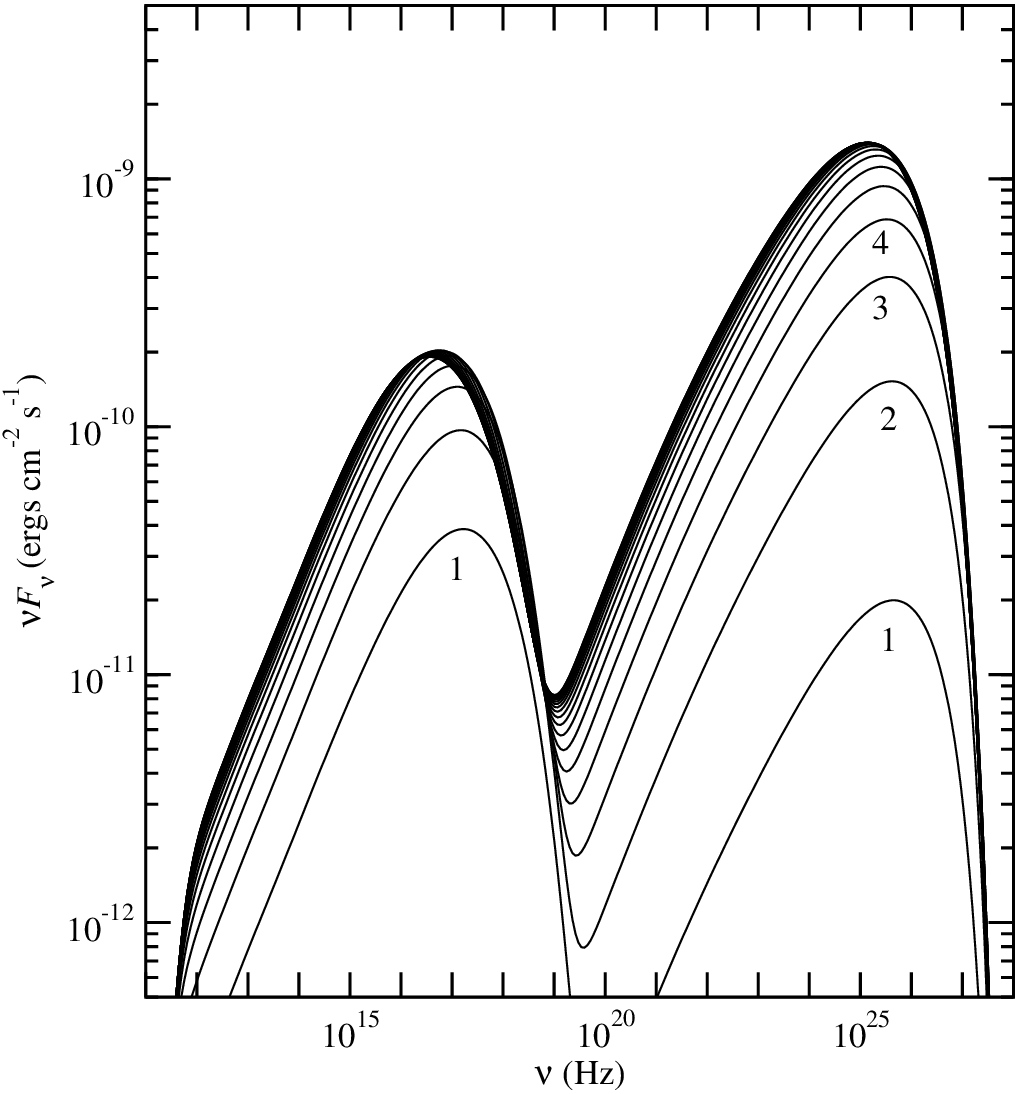}
\caption{Time evolution of SED of Model A is shown. The SEDs are plotted at every $R/c$
and evolve from lower to upper lines. Labels are $t/(R/c)$ in the comoving frame.}
\label{fig:photon-evolv}
\end{figure}

In our simulations, small numbers of nonthermal electrons and synchrotron photons are 
initially injected and time evolution is followed until a steady state is attained.
Because of this initial condition, the SED takes longer time than the observed timescale
to attain a steady state.
The time evolution of the SED of Model A is shown in Figure \ref{fig:photon-evolv}.
In actual flares, the initial condition may different from that used here
and flares are not in a steady state.
We used a steady state to obtain typical values of the source parameters.

\begin{figure}
\includegraphics[angle=0,scale=.59]{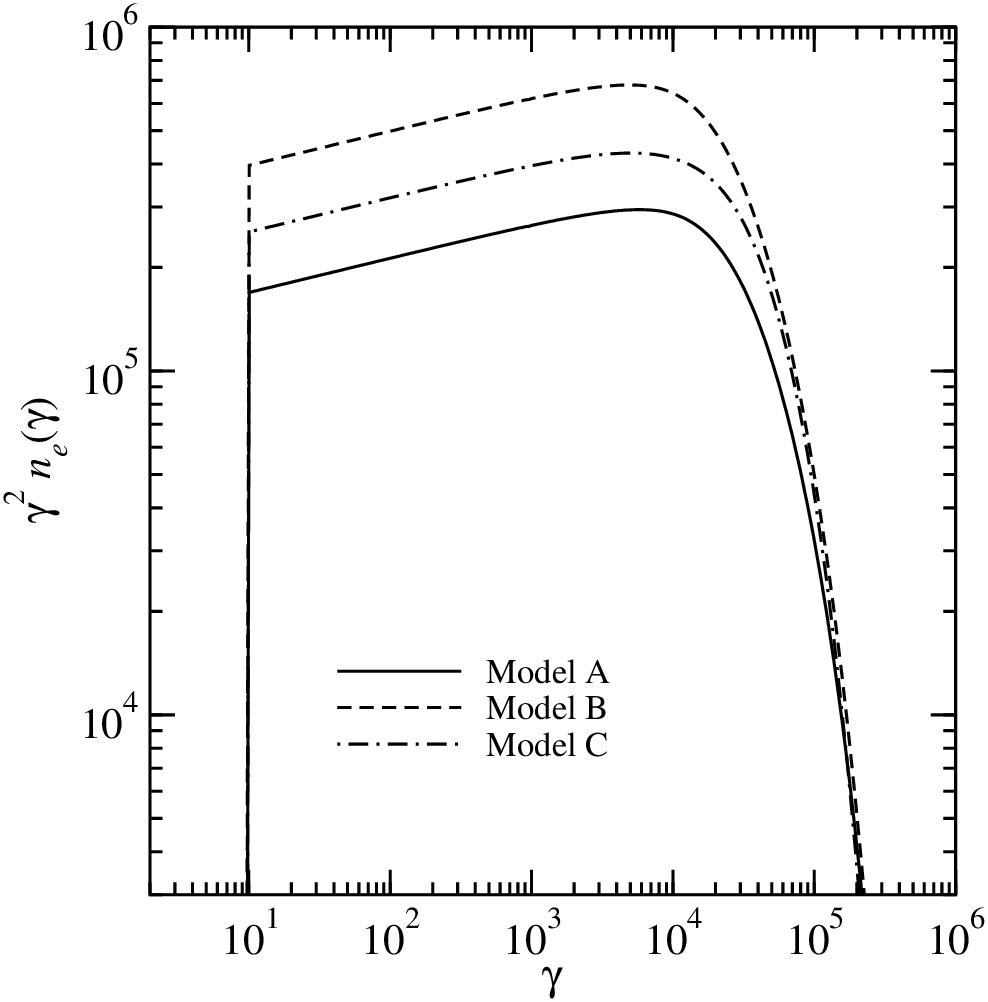}
\caption{The electron spectrum for the flare in 2006 July.
Models A, B, and C are shown by solid, dashed, and dot-dashed lines, respectively.}
\label{fig:electronspec1}
\end{figure}

In Figure \ref{fig:electronspec1} the electron spectrum for the flare is shown.
The electron spectrum has a cutoff at $\gamma \sim 10^4$.
Because the injection spectrum has an exponential cutoff at $\gamma_\mathrm{max} > 10^4$,
efficient Compton cooling made the cutoff energy smaller.
The synchrotron cooling time of electrons with $\gamma = 10^4$ is about 
$7.7 \times 10^6$ s for $B = 0.1$ G.
The energy density of synchrotron photons is 
$u_\mathrm{syn} \sim 5.3 \times 10^{-3}$ ergs cm$^{-3}$ in Model A.  
This results in the Compton cooling time $\sim$ $5.8 \times 10^5$ s 
for electrons with $\gamma = 10^4$.
Electrons with $\gamma \sim \gamma_\mathrm{max}$ are cooled in the flare timescale
in the comoving frame of the jet.
The Compton cooling time is short enough to be a major cooling process 
during the TeV flare.

The numerical results show that the energy densities of nonthermal electrons 
and magnetic fields in Model A are, respectively,
$u_e = 3.8$ ergs cm$^{-3}$ and $u_B = 4.0 \times 10^{-4}$ ergs cm$^{-3}$.
This yields a large value of $u_e/u_B \sim 9.5 \times 10^3$.
The values of energy contents of different models are shown in Table \ref{tbl-2}.
The dominance of particle kinetic energy over magnetic energy should be taken 
into account in consideration of acceleration mechanisms.
The powers contained in the jet are, in the form of radiation and electron kinetic energy, 
given by $L_\mathrm{rad} = \pi c R^2 \Gamma^2 u_\mathrm{rad}$ and
$L_\mathrm{kin} = \pi c R^2 \Gamma^2 u_e$, respectively,
where $u_\mathrm{rad}$ is the radiation energy density.
The numerical results are given in Table \ref{tbl-2}.

Our model include internal absorption of $\gamma$-rays by $e^\pm$ production.
However, our numerical results show that the $\gamma$-ray absorption is negligible.
Because of the large values of $\Gamma$, the soft photon density and the maximum $\gamma$-ray
energy in the comoving frame of the jet are not large enough for $e^\pm$ pair production.

In Figure \ref{fig:photonspec1} we show a TeV spectrum observed in 2003 July as well.
In 2003, TeV emission was observed by H.E.S.S. several times \citep{aha-quiet04}.
In those observations the TeV flux was much lower than that in the 2006 July flare.
The TeV spectrum of 2003 July are fitted with parameters such as
$\Gamma = {\cal D} = 40$, $B = 0.08$ G, $R = 2.5 \times 10^{16}$ cm, $p = 1.95$, 
$\gamma_\mathrm{min} = 10$, and $\gamma_\mathrm{max} = 2 \times 10^5$.
With these parameters, numerical results give $u_e/u_B = 26.7$.
Compared with the flare of 2006 July, a blob with a larger size 
but a smaller value of $\Gamma$ was involved.
The powers of radiation and electrons are, respectively,
$L_\mathrm{rad} = 1.7 \times 10^{43}$ ergs s$^{-1}$
and $L_\mathrm{kin} = 6.4 \times 10^{44}$ ergs s$^{-1}$.
Compared with the flare state in 2006, $L_\mathrm{kin}$ is smaller by a factor 0.24
and $L_\mathrm{rad}$ is smaller by a factor 0.52,
if Model A is adopted.

In \cite{aha-quiet-model05}, they derived source parameters for the H.E.S.S. observation 
in 2003, October and November, with different EBL models from that used here.
According to their results, the source was in a low or quiet state.
With their leptonic model, they obtained $R = 1.5 \times 10^{15}$ cm,
${\cal D} = 25$, and $B = 0.25$ G (Model 2).
\cite{ktk02}, on the other hand, used the SSC model to fit another quiescent state,
but without the correction for $\gamma$-ray absorption by EBL.
They obtained $R = 9 \times 10^{15}$ cm, ${\cal D} = 33$, and $B = 0.3$ G.
The ratio $u_e/u_B$ is 3, which is smaller by a factor 9, compared with our results
for 2003 July.  The source size obtained by us is the largest among these models
for quiescent states.


\section{Discussion}
\label{sec:sum}

In this work we have shown that a large value of the bulk Lorentz factor, 
$\Gamma \sim 100$, is required to explain the emission from PKS 2155--304
in the TeV band as well as the X-ray band.
With $R = 9.6 \times 10^{14}$ cm and $\Gamma = 90$, 
the timescale $t_\mathrm{var} \sim 400$ s is obtained,
while $t_\mathrm{var} \sim 200$ s is obtained for $R \sim 5 \times 10^{14}$ cm
and $\Gamma \sim  100$ - 150. 
The value of ${\cal D} = \Gamma$ as large as 100 requires the viewing angle
as small as 0.01 rad, which makes the case an unlikely coincidence 
if the opening angle of the jet is the same order as $1/\Gamma$.
However, if the opening angle is larger than $1/\Gamma$,
we can observe a portion of the jet that aligns with the line of sight,
thus avoiding an excessively small observability.
Considering that Mrk 501 also exhibited a similar short timescale TeV flare,
we think that the opening angle is larger than $1/ \Gamma$ (by a factor $a$).
If this is the case, $R$ represents the lateral size along the jet motion,
while the transverse size is larger than $R$ by the same factor $a$.
Thus the real size of the emission region is larger than $R$ 
and only a part of the emission regions is observed owing to a strong relativistic beaming effect.
Then the kinetic power should be by a factor $a^2$ larger than 
a conventional estimate given by $\pi c R^2 \Gamma^2 u_e$.
A further speculative  possibility is that the opening angle is
as small as $1/\Gamma$ but the jet axis wanders within a large angle 
and the flare is observed when its direction happens to fall in the line of sight.

Our results show that the jet plasma is highly particle energy dominated over 
magnetic energy, i.e., $u_e/u_B \sim 10^4$ in the flare state in 2006.
The ratio $u_e/u_B$ has been discussed in many papers. 
In \citet{ktk02} we discussed that TeV blazars are generally particle dominated
typically by a factor of 10.
For a quiescent state of PKS 2155--304 we obtained the ratio of 3.
Thus the ratio of $10^4$ seems unusually high,
although there is no serious theoretical reason to expect a ratio of unity
except that it leads to a minimum kinetic power for a given observed spectrum.
The obtained high value of the ratio basically results from the fact 
that the SSC luminosity is much larger than the synchrotron luminosity.
In this respect, we note that the X-ray and TeV observations are not strictly
simultaneous; if the X-ray flux during the TeV flare is higher than the adopted
value in this paper, the ratio decreases correspondingly.
Also, some soft photon sources other than synchrotron radiation will help to
decrease the ratio.  However, considering a good fit to the observed spectrum
obtained here, such possibilities are not easy to realize and the reduction will
not be significant.
\citet{lev07} and \citet{bgl08} adopted the view that the TeV flare is caused by
external Compton mechanism.  In this case, the model may avoid an extremely large
value of $u_e/u_B$.
However, considering the short timescale of radiative deceleration, 
the acceleration of the emitting plasma to the large bulk Lorentz factor itself
becomes very difficult because of the large radiation drag effect.
Also note that both papers did not make detailed multi-wavelength fitting.

As mentioned above, \cite{bgl08} argued that inverse Compton scattering off external soft photons 
is favored as the dominant radiation process to produce TeV $\gamma$-rays.
The major differences between their model and ours are that
they assumed the dominance of the Poynting flux over the kinetic
energy flux as well as the synchrotron radiation as a major cooling process 
to estimate particle's cooling timescale,
while in our model the energy flux is dominated by
nonthermal particle's kinetic energy and the cooling is dominated by SSC.

In comparison between the flare state in 2006 July and the quiescent state in 2003 July,
the powers in radiation and electrons increased by factors 2 and 4.2, respectively,
during the flare in 2006, if Model A is adopted.
The increase in the jet power is significant 
and the efficient acceleration of electrons is implied.
The size of the blob for the 2006 July flare is $\sim 10^{15}$ cm.
This is about 3 $R_\mathrm{S}$, if the mass of the central black hole is $10^9 M_\sun$.
A large energy injection into a very compact region occurred in the flare.
The obtained parameter values for the flare event and various phases are not well ordered
and we found no simple tendencies from quiescent states to flares.
The TeV flare may be caused by a concentration of energy into a sub-horizon scale region,
while during quiescent states we may be seeing larger scale energy dissipation on
longer timescales.

In applying our model, we assumed a low flux model of EBL, i.e., LLL of \cite{dk05}.
With this EBL model the one-zone SSC model is found to be viable for this TeV blazar,
although other models proposed by \cite{gk03} and \cite{gtc05} might explain the TeV 
spectrum with higher fluxes of EBL.  This should be studied in future work.

Radiation below $10^{15}$ Hz is not fitted with our model.
We assumed that the emission comes from more extended regions far away from the central
black hole \citep[see also][]{aha-quiet-model05}.
If the radiation below $10^{15}$ Hz is emitted by the jet plasma,
inverse Compton scattering produces radiation in the MeV -- GeV band.
Observation in this energy band is particularly important to set constraints
on the emission mechanisms of blazars.


\acknowledgments
This work has been partially supported by Scientific Research Grants 
(F. T.: 185402390) from 
the Ministry of Education, Culture, Sports, Science and Technology of Japan.



\clearpage

\clearpage

\begin{deluxetable}{cccccc}
\tablecaption{Model Parameters\label{tbl-1}}
\tablecolumns{6}
\tablewidth{0pt}
\tablehead{
\colhead{Parameter} & \colhead{Model A} &  \colhead{Model B} &  \colhead{Model C} & 
\colhead{Foschini et al. Model\tablenotemark{a}} & 
\colhead{Begelman et al. Model\tablenotemark{b}}
}
\startdata
$\Gamma$    & 90 & 100   & 120   & 30  & $\gtrsim 50$ \\
${\cal D}$         & 90  & 100   & 120  & 33.5 & $\cdots$ \\
$B$ (G)            & 0.1 & 0.14  & 0.08  & 0.27 & $> 1.6$ \\
$R$ ($10^{14}$cm)  & 9.6 & 5.4   & 6.5 &  50  & 4.5\\
$p$                & 1.9 &  1.9   &  1.9  & 2.5\tablenotemark{c} & $\cdots$ \\
$\gamma_\mathrm{max}$ ($10^4$) & 8.0 & 6.7 & 5.5  & 
17.5\tablenotemark{d}  & $\cdots$ \\
$\gamma_\mathrm{min}$ & 10 & 10 & 10 & $\cdots$ & $\cdots$ \\
$\gamma_\mathrm{break}$\tablenotemark{e} & $\cdots$ & $\cdots$ & $\cdots$ & 
$1.5 \times 10^4$  & $\cdots$  \\
$\gamma_\mathrm{peak}$ & $\cdots$ & $\cdots$ & $\cdots$ & $\cdots$ & $10^4$ \\
\enddata
\tablenotetext{a}{The model for July 29 given in \cite{fos07}. 
The variability timescale $\sim $1 h was assumed.}
\tablenotetext{b}{The model given in \cite{bgl08}.
$\Gamma = 50$, $t_\mathrm{var} = 300$ s, and isotropic luminosity 
$L_\mathrm{iso} = 10^{46}$ ergs s$^{-1}$ are
assumed to calculate $B$, $R$, and $\gamma_\mathrm{peak}$, where
$\gamma_\mathrm{peak}$ is the electron Lorentz factor
that emits synchrotron radiation peaking at $\sim 10^{16}/\Gamma$ Hz.}
\tablenotetext{c}{The power law index for $\gamma_\mathrm{min} < \gamma < \gamma_\mathrm{break}$.
The index for $\gamma_\mathrm{break} < \gamma < \gamma_\mathrm{max}$ is $p + 1$}.
\tablenotetext{d}{The maximum energy of electrons.}
\tablenotetext{e}{The break energy of electrons obeying a broken power law.}
\end{deluxetable}

\clearpage

\begin{deluxetable}{cccc}
\tablecaption{Energy Contents \label{tbl-2}}
\tablecolumns{4}
\tablewidth{0pt}
\tablehead{
\colhead{Parameter} & \colhead{Model A} &  \colhead{Model B} &  \colhead{Model C} 
}
\startdata
$\Gamma$   & 90  & 100   & 120 \\
$u_e$ (ergs cm$^{-3}$) & 3.8  &  8.7  &  5.7  \\
$u_B$ ($10^{-4}$ ergs cm$^{-3}$) & 4.0 & 7.8  &  2.6  \\
$u_e/u_B$ ($10^4$)  & 0.95 &  1.1  & 2.2  \\
$L_\mathrm{rad}$ ($10^{43}$ ergs s$^{-1}$) & 3.3 & 2.9  & 1.7 \\
$L_\mathrm{kin}$ ($10^{45}$ ergs s$^{-1}$) & 2.7 & 2.4  & 3.3  \\
\enddata
\end{deluxetable}


\begin{thebibliography}{}
\bibitem[Aharonian et al. (2005a)]{aha-quiet04}
Aharonian, F. A., et al. 2005a, \aap, 430, 865

\bibitem[Aharonian et al. (2005b)]{aha-quiet-model05}
Aharonian, F. A., et al. 2005b, \aap, 442, 895

\bibitem[Aharonian et al. (2007)]{aha-flare07}
Aharonian, F. A., et al. 2007, \apj, 664, L71

\bibitem[Albert et al. (2007)]{alb07}
Albert, J. et al. 2007, \apj, 669, 862

\bibitem[Begelman et al. (2008)]{bgl08}
Begelman, M. C., Fabian, A. C., \& Rees, M. J. 2008, \mnras, 384, L19

\bibitem[Dermer \& Schlickeiser (1993)]{ds93}
Dermer, C. D., \& Schlickeiser, R. 1993, \apj, 416, 458

\bibitem[Dwek \& Krennrich (2005)]{dk05}
Dwek, E., \& Krennrich, F. 2005, \apj, 618, 657

\bibitem[Foschini et al. (2007a)]{fos07}
Foschini, L., et al. 2007a, \apj, 657, L81

\bibitem[Foschini et al. (2007b)]{fos-ast07}
Foschini, L., et al. 2007b, in AIP Conf. Proc. 921, The First Glast Symposium, 
ed. S. Ritz \& P. Michelson (Melville: AIP), 329

\bibitem[Georganopoulos \& Kazanas (2003)]{gk03}
Georganopoulos, M., \& Kazanas, D. 2003, \apj, 594, L27

\bibitem[Ghisellini et al. (2005)]{gtc05}
Ghisellini, G., Tavecchio, F., \& Chiaberge, M. 2005, \aap, 432, 401

\bibitem[Gould \& Schr{\'e}der (1966)]{gs66}
Gould, R. J., \& Schr{\'e}der, G. 1966, Phys. Rev. Lett., 16, 252

\bibitem[Jelly (1966)]{jelly66}
Jelly, J. V. 1966, Phys. Rev. Lett., 16, 475

\bibitem[Kato et al. (2006)]{kkt06}
Kato, T., Kusunose, M., \& Takahara, F. 2006, \apj, 638, 658

\bibitem[Kino et al. (2002)]{ktk02}
Kino, M., Takahara, F., \& Kusunose, M. 2002, \apj, 564, 97

\bibitem[Kusunose et al. (2000)]{ktl00}
Kusunose, M., Takahara, F., Li, H., \apj, 536, 299

\bibitem[Levinson (2007)]{lev07}
Levinson, A. 2007, \apjl, 671, L29

\bibitem[Maraschi et al. (1992)]{mar92}
Maraschi, L., Ghisellini, G., \& Celotti, A. 1992, \apjl, 397, L5

\bibitem[Mannheim \& Biermann (1992)]{mb92}
Mannheim, K., \& Biermann, P. L. 1992, \aap, 253, L21

\bibitem[M\"{u}cke et al. (2003)]{mu03}
M\"{u}cke, A. et al. 2003, AstroPart. Phys., 18, 593

\bibitem[Nikishov (1962)]{nik62}
Nikishov, A. I. 1962, Sov. Phys. J. Exp. Theor. Phys. 14, 393

\bibitem[Piner \& Edwards (2004)]{pe04}
Piner, B. G., \& Edwards, P. G. 2004, \apj, 600, 115

\bibitem[Sikora et al. (1994)]{sikora94}
Sikora, M. et al. 1994, \apj, 421, 153

\bibitem[Stecker et al. (1992)]{sjs92}
Stecker, F. W., de Jager, O. C., \& Salamon, M. H. 1992, \apjl, 390, L49

\bibitem[Stecker et al. (2006)]{sms06}
Stecker, F. W., Malkan, M. A., \& Scully, S. T. 2006, \apj, 648, 774 
[Erratum \apj, 658, 1392, 2007]

\bibitem[Wagner (2007)]{wag07}
Wagner, R. M. 2008, MNRAS,386, 199
\end{thebibliography}
\end{document}